\documentclass[journal]{IEEEtran}

\usepackage{cite}
\usepackage{array} 
\usepackage{multirow}
\usepackage{amsmath,amssymb,amsfonts}
\usepackage{algorithm2e}
\usepackage{graphicx}
\usepackage{textcomp}
\usepackage{subcaption}
\usepackage{xcolor}
\usepackage{amssymb}
\def\BibTeX{{\rm B\kern-.05em{\sc i\kern-.025em b}\kern-.08em
    T\kern-.1667em\lower.7ex\hbox{E}\kern-.125emX}}
\usepackage{soul}

\begin{document}

\title{EvoCreeper: Automated Black-Box Model Generation for Smart TV Applications}

\author{Bestoun S. Ahmed, and 
        Miroslav Bures%
\thanks{This study is conducted as a part of the project TACR TH02010296 ``Quality Assurance for Internet of Things Technology''.}
\thanks{B. Ahmed is with the Department of Mathematics and Computer Science, Karlstad University, Sweden and the Department of Computer Science, Czech Technical University, Karlovo nam. 13, Prague, Czech Republic,
email: bestoun@kau.se}%

\thanks{M. Bures are with the Department of Computer Science, Faculty of Electrical Engineering, Czech Technical University, Karlovo nam. 13, Prague, Czech Republic}%

}

\maketitle

\begin{abstract}
Smart TVs are coming to dominate the television market. This accompanied by an increase in the use of the smart TV applications (apps). Due to the increasing demand, developers need modeling techniques to analyze these apps and assess their comprehensiveness, completeness, and quality. In this paper, we present an automated strategy for generating models of smart TV apps based on a black-box reverse engineering. The strategy can be used to cumulatively construct a model for a given app by exploring the user interface in a manner consistent with the use of a remote control device and extracting the runtime information. The strategy is based on capturing the states of the user interface to create a model during runtime without any knowledge of the internal structure of the app. We have implemented our strategy in a tool called EvoCreeper. The evaluation results show that our strategy can automatically generate unique states and a comprehensive model that represents the real user interactions with an app using a remote control device. The models thus generated can be used to assess the quality and completeness of smart TV apps in various contexts, such as the control of other consumer electronics in smart houses.

\end{abstract}


\IEEEpeerreviewmaketitle

\section{Introduction}
The smart TV is a modern technological device that is a hybrid of a computer and a traditional television. In addition to a conventional TV terminal, this device incorporates digital content and an operating system (OS) with an Internet connection. Smart TVs usually provide access to broadcast media, games, digital services, various online interactive sessions, on-demand entertainment, Internet browsing, and many other services, and these devices are expected to become even more intelligent, interactive, and useful in the future \cite{Jung2011}. Investments in related technological advancements by electronics companies and IT firms have recently been increasing. As a result, new terminals and applications for smart TVs have been launched. It is expected that these devices will soon become a common feature of smart homes within an Internet of Things (IoT) context \cite{BussinessInside2016}. This explains why the smart TV market had grown to be worth \$265 billion by 2016 \cite{MarketsandMarkets}. 

Like all new smart devices, a smart TV is operated by an OS that handles the necessary hardware interaction functionality and a set of applications (apps) installed on the OS to provide various services to the user. Despite the visual similarities between smart TV apps and mobile apps, the mode of user interaction with smart TV apps is different. For mobile apps, the user interacts with the touchscreen of the device (i.e., the apps) directly by hand, whereas for smart TVs, the user interacts with an app through another device, namely, the remote controller. Of course, some vendors also provide touchscreen interactions to users. Additionally, some recent studies have investigated gaze-based interactive interface design for smart TV apps \cite{Chen2018}. However, the way in which a smart TV behaves is still primarily based on a remote control device when it comes to the navigation of the user interface (UI) states. Moreover, the user of any TV (including smart TVs) is usually far away from the screen and uses the remote controller to operate apps almost all the time. 

With the increasing number of smart TV apps, there is an urgent need for methods of modeling UIs based on user interactions. As in the case of the UIs of desktop or mobile apps, such a model can be used for many purposes. It can be used to assess the completeness of an app during development or to assess the quality of the app during testing. Such a model can also be used to trigger requirement specifications and analyze poorly documented legacy apps. It can also be used in the smoke testing of apps. In fact, most developers currently create mental models of UIs in order to better comprehend the software.

Creating a model of the UI via the user interaction mode is a common first step in assessing the quality and completeness of a UI-based app. While model generation is a common step, the presentation of the model depends mainly on the type of app and the user interaction mode. For a smart TV app, if the app is to be used on a touchscreen TV, the same state-of-the-art model generation methods used for mobile apps may also be useful when deployed in the smart TV operation environment. However, there is still a need to design a model generation method for gaze interaction with an app, and our method may not be useful for this purpose. Thus, in general, the model of a smart TV app does not look like the model of a mobile app due to the different mechanism of transitioning among the UI states. In a mobile app, a user can easily transition among app states by going directly from one state to another. However, for example, in a smart TV app, when the user wants to go from one state to another, he must pass through some set of states by means of the remote control device in order to reach the destination state. 

It is possible to create a model using the same state-of-the-art approaches used for the reverse engineering of mobile apps. In practice, however, that model will be useless for a smart TV app. Normally, a model is created for some specific purpose. For example, we may create a model to generate several test cases. However, it will be impossible to run those test cases on a real smart TV because there will be significant differences in the transitions among the states of the UI.

In this paper, we propose a strategy called EvoCreeper\footnote{EvoCreeper can be downloaded here: https://bit.ly/2StQeG4} for the automatic creation of comprehensible UI models for smart TV apps. EvoCreeper is a generic open-source model generator for smart TV apps that can generate directed graphs by reverse engineering an app without knowledge of its internal code structure. In this strategy, the UI states and the transitions among them are examined in order to create the model. The contributions of this paper can be summarized as follows:

\begin{itemize}
\item We present a new comprehensive automated black-box strategy in which a smart TV app is executed on an emulator and information is extracted from the UI during runtime.  

\item We present an efficient exploratory algorithm that can explore the elements of a UI by simulating user interactions with a smart TV app.  

\item We illustrate the implementation of our strategy within our EvoCreeper tool, which can create directed graph models of smart TV apps. The models thus generated can be used for various development and testing purposes. 

\item We have developed an algorithm to verify the correctness of the generated models. 

\item We have empirically evaluated our strategy through four  real case studies. 
\end{itemize}

The rest of this paper is organized as follows. In Section \ref{relatedWork}, we summarize the technological background of smart TV apps. This section also mentions the related challenges and the most closely related works in the literature. Section \ref{ourApproach} summarizes the details of our approach for generating models of smart TV apps. Section \ref{fig:Proof-of-concepts} presents a proof of concept for our model generation approach. Section \ref{Evaluation} reports the evaluation results, and Section \ref{Threats} summarizes the threats to the validity of our evaluation experiment. Finally, Section \ref{Conclusion} gives concluding remarks and discusses our future research directions. 

\section{Background and Related Work}\label{relatedWork}

Smart TV apps are developed using software development kits (SDKs). Each platform has its own SDK for the development of software for TV devices. For example, the Android and Tizen SDKs can be used for smart TV app development. Recently, a few SDKs have also begun to support cross-platform development. For instance, the Mautilus \cite{mautilusSDK} Smart TV SDK is an example development framework, but at present, the apps developed in this framework work on only some versions of the supported devices. The Smart TV Alliance \cite{SmartTVAlliance} was another project for supporting cross-platform development; however, this project has been inactive for some time. In fact, the Tizen SDK is currently the SDK that is most commonly used since it provides a set of tools and frameworks for the development of smart TV apps through Tizen Studio that utilize the latest web technologies, such as JavaScript, CSS, HTML5, and W3C widget packaging, which are used by most smart TV apps. Additionally, JavaScript is used in most apps as a standard programming language for programming their behavior. The use of JavaScript endows an app with the page jumping capability. It also enables the developer to code complex expressions and calculation structures such as conditional branches and loops.

In general, a smart TV app can be of one of two types: installed or cloud-based. An installed TV app is a stand-alone app installed on a smart TV without the need for an Internet connection, while a cloud-based TV app primarily acts as an interface between the cloud and the TV and offers only shallow content (almost no additional functionality) when there is no Internet connection available.

Regardless of the visual appearance of these apps, mobile and smart TV apps are different in several significant respects. For example, the size of the screen can affect the layout of an app. Smart TVs have wider screens than small mobile devices do. The background color of a smart TV app may also be different from the corresponding color on mobile devices. The size of the icons could also be different. From the perspective user interaction, smart TV apps typically involve less text entry because of the difficulty of entering text using a remote control device. Most smart TV apps are designed to retrieve content from the Internet, whereas this is not the case for mobile apps, which can be standalone apps without Internet connection interfaces \cite{OperahDesignDevelope}. The typical smart TV app is much more straightforward than the typical mobile app, especially in its design layout. We have explained these analogies and differences in detail in our previous study \cite{BestounFTCReview}.

The way in which the user interacts with the app constitutes an essential difference between smart TV and mobile apps. The user of a mobile app interacts directly with the app without an intermediate device, while for a smart TV app, the user interacts with the help of a remote controller. In fact, the UIs of smart TV apps are sometimes called 10-foot UIs since 10 feet (3 m) is the standard distance between the user and the TV. Developers consider this distance when developing smart TV UIs \cite{Sabina2016,OperahDesignDevelope}. Using a remote control device at this distance is not a user-friendly or responsive experience. Hence, the design of a smart TV UI must consider this significant difficulty.

Navigation in a smart TV app is achieved through a remote control device. Although some new TV devices offer the ability for the user to directly interact with the screen, the most common form of interaction with a TV device is still through a remote control device. A remote control device includes four essential navigation buttons: \textit{Right} $\Rightarrow$, \textit{Left} $\Leftarrow$, \textit{Up} $\Uparrow$ and \textit{Down} $\Downarrow$. Additionally, a remote control device has an \textit{OK} button to choose any selected item in an app after navigating to it and a \textit{Back} button $\hookleftarrow$ to navigate back to the previous screen. These six key buttons should work properly when using an app.

In addition to these six buttons, there are many other buttons on remote control devices that vary from one TV brand to another depending on the level of the functionalities they access. Some of them are related to the hardware functionalities of the TV itself. For example, the power button turns the TV on and off. There are typically also ten number buttons (from 0 to 9) for channel jumps and entering numbers in text fields if necessary.

\begin{figure*}
\begin{centering}
\includegraphics[scale=0.65]{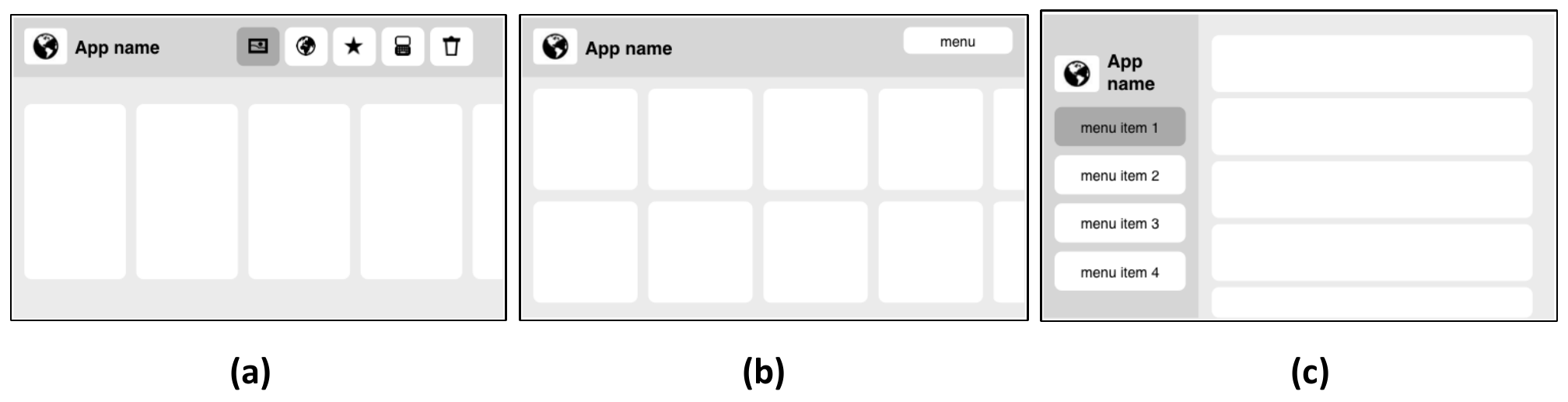}
\par\end{centering}
\caption{\label{fig:Three-main-layout}Three main layout design patterns for smart TV apps \cite{OperahDesignDevelope}}
\end{figure*}

The UI layout of any app plays a primary role in (black-box) model construction. A better understanding of this layout can lead to a more accurate model. Smart TV apps typically follow one of a limited number of layout patterns. Figure \ref{fig:Three-main-layout} shows the three main patterns followed by most smart TV apps. Of these, layout (b) is most commonly used since it presents many items on a single screen.

\begin{figure*}
\begin{centering}
\includegraphics[width= 2.8 in]					{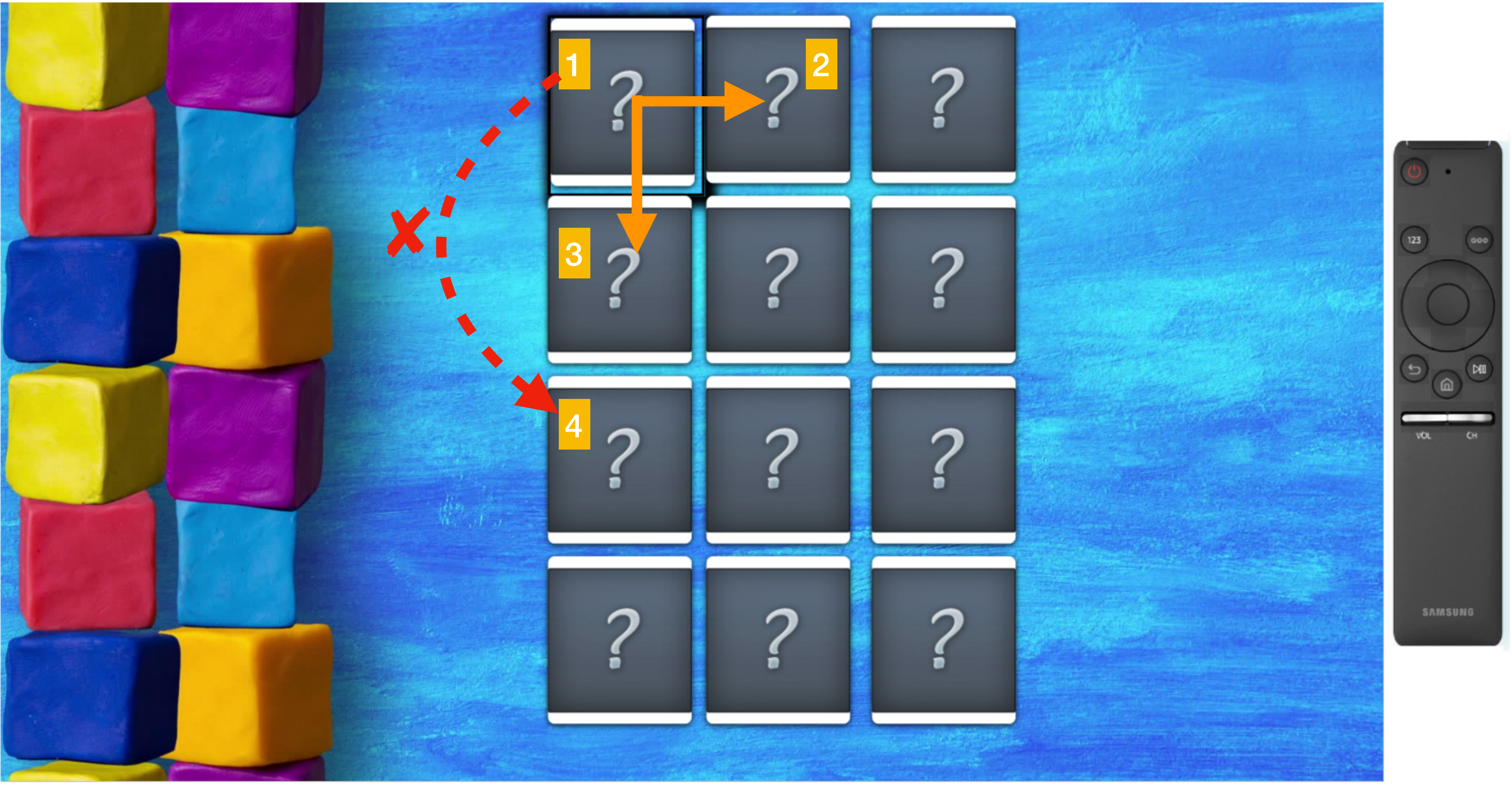}
\par\end{centering}
\caption{\label{fig:AppNavigation} A puzzle game smart TV app to illustrate the valid and invalid move based on the remote device}
\end{figure*}

The remote control device places constraints on the navigation from one state to another because it supports only one step of navigation at a time. Hence, each move in the layout is a step. Accordingly, the transition from one state to another is not smooth as in mobile or desktop apps. To move from one state to another in a smart TV UI, the user may need to pass through several other states before reaching the desired state. For nonadjacent states, more than one step is required to move from one to the other. As a simple example, we consider the UI of a puzzle game smart TV app, as shown in Figure \ref{fig:AppNavigation}. 

Clearly, there are 12 items (each regarded as a state) in the UI of the app in Figure \ref{fig:AppNavigation}. For clearer illustration, we have numbered these states. For instance, state 1 is the starting state. From here, there are only two valid moves: \textit{Right} to state 2 or \textit{Down} to state 3. Note that the user cannot jump directly from state 1 to state 4; we consider this to be an invalid move (red arrow). Hence, to move from state 1 to state 4, the user must press \textit{Down} twice, passing through state 3.

Generally, model generation for UI-based software apps is performed frequently for many purposes during the development process, as described in the literature. Such models have been used in the literature for testing and quality assurance purposes (e.g., \cite{Yuan:2011:GIT,Saddler:2017,Long:2013:NTM,Mehlitz:2011,AHMED2014218}). Regardless of the app type, as long as the model is for the UI, it must simulate the user’s interaction with the app in some form. Using this approach, Memon \textit{et al.} \cite{MemonReverse2003,Nguyen2014:Guitar} proposed a reverse engineering technique called GUITAR for modeling the UIs of desktop apps. The technique starts from the main window and automatically captures UI widgets to construct an event-flow graph model. That model is then used to generate test cases for program testing. This technique has undergone considerable development in many subsequent research papers. Aho \textit{et al.} \cite{AHO201449} presented an extensive survey of those studies and techniques. This approach has also been used for model generation for mobile apps. For example, Joorabchi and Mesbah proposed a reverse engineering technique for generating models of mobile apps, and Amalfitano \textit{et al.} \cite{Amalfitano2015MobiGuitar} developed the MobiGUITAR strategy for creating models of mobile apps.

Mesbah \textit{et al.} \cite{Mesbah:2012} also proposed a reverse engineering technique called CRAWLJAX for generating models of web apps. However, this technique relies on a dynamic crawler acting on a web app and detecting the clickable states. The model thus generated can then be used to generate test suites and analyze the app. In fact, this technique works like a gray-box technique rather than a black-box technique because it also scans the code of the app.

Gimblett \textit{et al.} \cite{Gimblett:2010} tried to define a generic approach for establishing models of interactive software to simulate user actions. It is clear from the literature that most related techniques follow the same basic principles. However, the technology, application type, and user interaction mechanism introduce differences and thus pose challenges for model generation.

As another class of smart devices, smart TVs are currently becoming increasingly popular due to the rapid development of apps that can be installed on these devices and the ability to control other consumer electronics and connect with them in the IoT context. Some examples of the possible applications of such technology include controlling home appliances through smart TVs\cite{Kim2013,Kim2015}, home sleep care with video analysis using a smart TV app \cite{Fan2014}, controlling smart homes from smart TVs \cite{Cabrer2006}, healthcare applications for smart homes \cite{Vavilov2014}, smart lighting control \cite{Chun2013}, and smart security camera systems \cite{Erkan2015}. With the increasing prevalence of these applications, it is becoming increasingly difficult to ignore the unique requirements of quality assessment for smart TV apps and assume that the same quality assessment procedures can be used as for mobile apps. As with other smart devices, creating a model of a smart TV app is the first step towards quality assessment.

To create a model for a smart TV app, it is necessary to detect the active states of the app and the transitions among them. While the same basic principles and concepts are followed as in previous approaches to detect new states and state transitions, as mentioned previously, there are significant differences due to the different technologies and modes of user interaction. In desktop GUIs and web apps, the combination of a keyboard and mouse is still the standard mode of user input for interacting with these apps. However, this is not the case for mobile apps because the user interacts with the touchscreen of the device by means of his or her fingers, and hence, different users will exhibit different interaction behaviors. Although this issue has led to the development of new models for mobile apps, many of these strategies still benefit, wholly or partially, from the earlier methods and practices established for the reverse engineering of desktop and web apps. Nevertheless, the differences in interaction create many obstacles and difficulties. For example, Nguyen \textit{et al.} \cite{Nguyen2014:Guitar} used an event-flow graph (EFG) as a model of the UI of a desktop app, whereas Amalfitano \textit{et al.} \cite{Amalfitano2015MobiGuitar} used a state machine as a model for a mobile app due to the different natures of their interactions. For smart TV apps, neither EFG nor state machine models are applicable. In an app of this kind, each transition from one state to another is, in practice, just one step, while this is not the case in other apps. For example, in a mobile app, the distance between two icons (states) is irrelevant to the transition, whereas this is a critical issue in a smart TV app, and this difference will lead to a different model.

A significant effort to formulate such a model has been made recently by Cui \textit{et al.} \cite{Cui2017}. In their study, a hierarchical state transition matrix (HSTM) was proposed as a model for an Android smart TV app. While this model is promising, there is a need to further develop and formulate it for the complex structures of different apps. However, this type of model is useful when model optimization and reduction are needed. In fact, Cui \textit{et al.} \cite{Cui2017} used a white-box crawler approach to scan the code of an app to construct a preliminary model. This initial model contains many obsolete nodes because the crawler detects all views present in the UI code. Even when a view is not an active state in the UI, the crawler algorithm considers it as a node in the model. For example, a piece of text in an Android app is a view; however, in practice, it is not a clickable state. When this approach is adopted, there is a possibility of combinatorial explosion in the resulting model. Therefore, the authors proposed an algorithm for reducing and optimizing the model by distinguishing obsolete and active nodes. However, this could be a time-consuming task and may lead to deviation from the primary research focus while also adding overhead to the model construction process.

In contrast to the contribution of Cui \textit{et al.} \cite{Cui2017}, our approach does not scan the app code. Our EvoCreeper strategy instead explores the UI of a smart TV app by examining each element and observing its reaction. When an element is clickable, the strategy will consider it as a state in the model. Section \ref{ourApproach} presents the details of our strategy.

\section{Our Model Construction Strategy}\label{ourApproach}

In this section, we present our new strategy for automatically generating a model of a smart TV app. Our strategy has been implemented with the Tizen SDK, which includes a smart TV emulator; however, the proposed framework is a general one, and it can be applied in combination with other possible emerging SDKs in the future. The strategy depends on the black-box approach to model generation and does not require knowledge of the internal structure of the app code.

The smart TV app is modeled as a directed graph $G=(N, E)$, where $N$ is a set of nodes, $N\neq\emptyset$, and $E$ is a set of edges. $E$ is a subset of $N\times N$ possibilities. In the model, we define one starting node, $n_{s}\in N$. The set $N_{e}\subseteq N$ contains the end nodes of the graph, where $N_{e}\neq\emptyset$. Each node corresponds to a UI element (state) of the app. Each edge corresponds to a possible transition between the states focused on by the cursor. These transitions can be triggered by individual keys on the remote control device. In this study, $e\in\{ D_{up}, D_{down},D_{left}, D_{right}, D_{OK}\}$ for each $e\in E$. 

If the model allows parallel edges, formally being a directed multigraph $G'=(N,E,n_{s},N_{e},s,t)$, such that $N\neq\emptyset$ is a finite set of nodes, $E$ is a set of edges, $s:E\rightarrow N$ assigns each edge to its source node and $t:E\rightarrow N$ assigns each edge to its target node. The node $n_{s}\in N$ is the initial/start node of the graph $G$ and $N_{e}=$ \{$n_{e}\mid n_{e}\in N$ has no outgoing edge \} defines nonempty set of end nodes of graph $G'$. In this study, approach we decided to base the app model on $G$, which practically does not allow parallel edges, to have the proposal in accord with the current path-based testing approaches, for instance, \cite{offutt2016introduction,li2012better,dwarakanath2014minimum}.

To detect all the necessary states in the app UI for presentation in the model, we have developed an algorithm called EvoCreeper. At present, the concept of state detectors for the UIs of mobile, desktop, and web apps is rather familiar. However, such detectors differ for each app category. As mentioned earlier, algorithms called crawlers have been developed that can crawl a UI and detect states. Because we are following a new approach to state detection, we do not call our algorithm a crawler. Rather, from a linguistic point of view, the name "creeper" is perfectly suited to what the algorithm does, whereas the word "crawler" carries a different meaning due to its use in web and search engine technologies. Algorithm \ref{alg:Application-Creeper} shows the steps of the EvoCreeper algorithm.

\begin{algorithm}

\KwIn {$v_{1}$ is the starting or user selected states}

\KwOut {List of states to be modeled $L_{v}$}

Iteration $It\leftarrow$1

Maximum Iteration $It_{max}\leftarrow max$

\While{$((It<It_{max})\parallel(newView\neq null))$}{

Use $v_{1}$ as a start point

From $v_{1}$ generate five possible directions $D_{Up}$ , $D_{Down}$ , $D_{Left}$ , $D_{Right}$, and $D_{OK}$

\ForEach {direction $D$ }{

Navigate a step

Monitor emulator log for reaction

\If {$newState=Active$}{

\If {the state is not duplicate} {
Add newState to $L_{v}$

\Else {Back to the parental state}

Record the in/out transitions
}
}
{\small{}$\quad$}$It++$

}

}

\caption{\label{alg:Application-Creeper} EvoCreeper Steps }

\end{algorithm}

One of the problems that EvoCreeper encounters when exploring a UI is establishing the position of the navigational cursor. Technically speaking, from the perspective of a JavaScript developer, this problem arises when a focus point is not set in the app. For several apps on the store, no focus point has been set by the developers. As a result, when such an app runs on the emulator, there are no preselected states in its UI. Instead, the user must use the remote control device to choose a state. Hence, a starting point for the navigator is missing. This problem is common with cloud-based smart TV apps because the UI changes in real time with the cloud content. Therefore, our strategy starts by checking the initial cursor condition. If a focus point is not set in the app, the strategy starts by asking the user to choose at least one state in the UI from which to start. From this state, the creeper will start creeping the UI evolutionarily and incrementally. If a focus point is already set, the strategy will proceed without asking the user for input.

The algorithm has five directions {\small{}$D_{Up}$, $D_{Down}$, $D_{Left}$, $D_{Right}$, and $D_{OK}$ } in which it can move from each state. When a new state is discovered in each direction (i.e., {$newState=Active$}), the algorithm will add it to the list of states to be modeled, $L_{v}$. The algorithm will continue until no new states are discovered. At this stage, the algorithm will choose to move back to the parent state. As an alternative stopping criterion, the algorithm will perform some preset number of iterations to avoid the possibility of an endless discovery loop that is encountered in some special cases of cloud-based apps.

In the post-processing phase, states $L_{v}$ are converted to nodes of the app model $N$. In this phase of the strategy, we used the direct mapping between the states and $N$, practically speaking, for each $l \in L_{v}$, an original $n \in N$ is created.

\begin{figure*}
\centering
\includegraphics[width=6 in]					{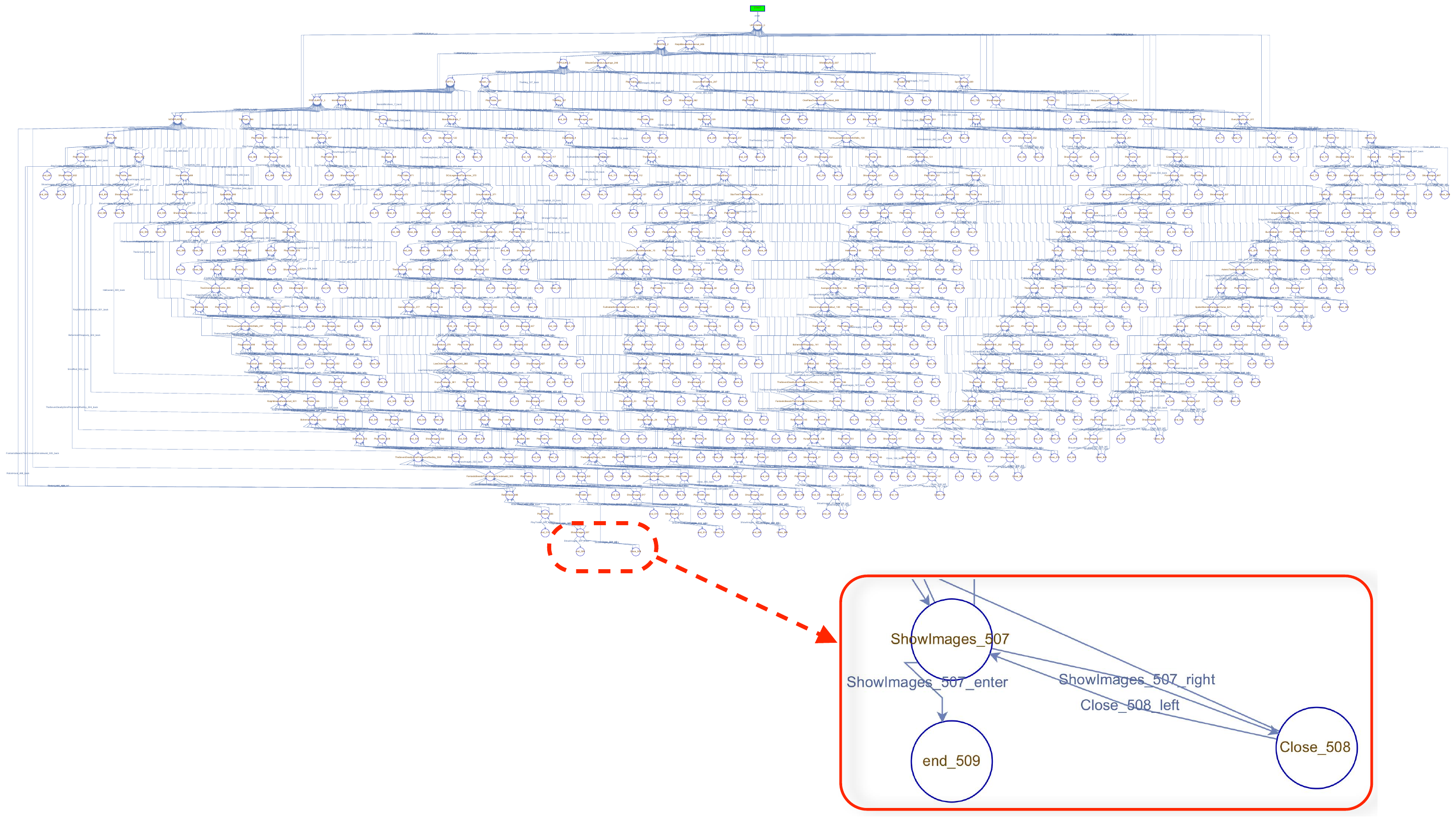}
\caption{The constructed directed graph model by EvoCreeper for the  CineMup}
\label{Figure:ExecutionTimeBox}
\end{figure*}

Figure \ref{Figure:ExecutionTimeBox} shows the directed graph model constructed by our strategy for the CineMup smart TV app. To illustrate the detail of the generated model, Figure \ref{Figure:ExecutionTimeBox} also shows a snapshot of three nodes in the graph. Here, each node has a unique identifier number preceded by the actual name of the state on the smart TV app. The model also records the actual transitions that can be performed from a state to another. Each transition name is preceded by the name of the state that is originated from.

The app model $G$ can be subsequently used for two principal purposes: (1) model checking, where a potential design sub-optimalities can be detected, and (2) automated path-based test case generation, being the major expected use case. Here, the goal is to generate a set of test cases $T$ that satisfy a defined test coverage criteria and counted optimal by a test set optimality criteria.

Respecting the standard approaches used in the field, a test case $t \in T$ is a sequence of nodes $n_{1},n_{2},..,n_{n}$, with a sequence of edges $e_{1},e{}_{2},..,e_{n-1}$, where $e_{i}=(n_{i,}n_{i+1})$,
$e_{i}\in E$. The test case $t$ starts with the start node $n_{s}$ ($n_{1}=n_{s}$) and ends with a $G$ end node ($n_{n}\in N_{e}$) \cite{offutt2016introduction,li2012better}. Test coverage criteria determine the strengths of the $T$ in the sense of number of alternative paths that are toured in the app during the tests and vary from low test coverage levels (node coverage or edge coverage) to very intense levels as all paths coverage \cite{offutt2016introduction}. For the test set optimality criteria, several options can be employed, for instance, the number of test cases, $|T|$, average length of the test cases,  ${\displaystyle \overline{|t|}={\frac {1}{|T|}}\sum _{i=1}^{|T|}|t_{i}|}, t_i \in T$, or total length of a test set ${\displaystyle l=\sum _{i=1}^{|T|}|t_{i}|}, t_i \in T$ \cite{li2012better,dwarakanath2014minimum}.
Such automated generation of test cases, which is subject of other studies, for instance, \cite{li2012better,dwarakanath2014minimum,sayyari2015automated,bures2017prioritized,arora2017synthesizing} open variety of options for construction of automated test frameworks for Smart TV apps and document the applicability potential of our strategy.

In the following section, we present an example as a graphical proof of concept for Algorithm \ref{alg:Application-Creeper}.

\section{Proof of Concept\label{subsec:Proof-of-Concept}}

\begin{figure*}
\begin{centering}
\includegraphics[scale=0.32]{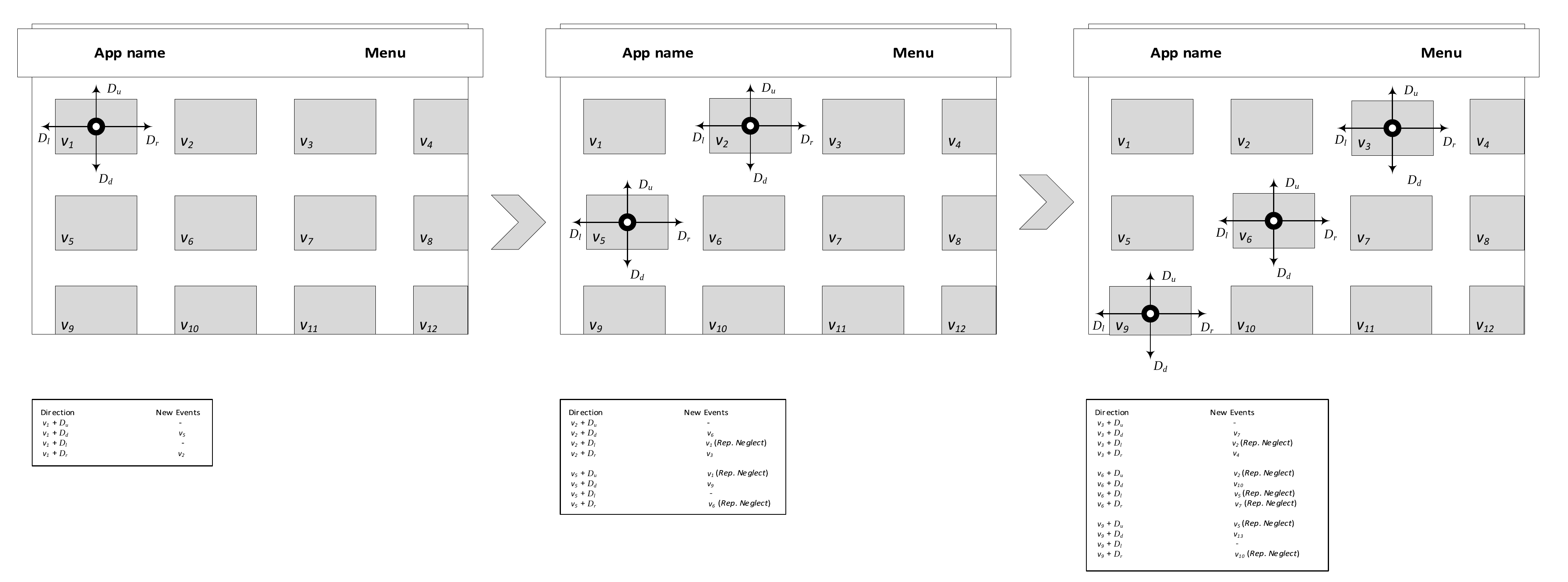}
\par\end{centering}
\caption{\label{fig:Proof-of-concepts}Proof of concepts of the EvoCreeper }
\end{figure*}

In this section, we present a proof of concept for the EvoCreeper concept introduced in Algorithm \ref{alg:Application-Creeper}. Here, we consider a cloud-based app as a pilot example because this is the most difficult scenario. As shown in Figure \ref{fig:Proof-of-concepts}, each active window has 12 states, and when the user shifts down or to the right, new states may appear. We consider three iterations of the algorithm. We assume that the user will choose $v_{1}$ as the starting state. In fact, $v_{1}$ is the worst-case choice among the states; we observe that choosing the state in the middle of the window instead may lead to fewer iterations and better recognition of the states. From $v_{1}$, the algorithm will consider four main directions: $D_{Up}$, $D_{Down}$, $D_{Left}$, and $D_{Right}$. For this proof of concept, we do not consider $D_{OK}$ because we are interested in the exploration of the current window, whereas $D_{OK}$ will probably take the algorithm to another window.

In each direction, the creeper algorithm will check for new states, which are most likely new elements in the UI. Considering the first iteration and starting from $v_{1}$, the up and left directions, $D_{u}$ and $D_{l}$, do not lead to new states, while the right direction, $D_{r}$, leads to $v_{2}$, and the down direction, $D_{d}$, leads to $v_{5}$. In the next iteration, the algorithm will start from the newly discovered states (here, $v_{2}$ and $v_{5}$). From $v_{2}$, the new states $v_{3}$ and $v_{6}$ are identified by the algorithm. In addition, $v_{1}$ is discovered in the $D_{l}$ direction; however, this state is ignored by the algorithm since it is already included in the state list. From $v_{5}$, the states $v_{1}$, $v_{9}$, and $v_{6}$ are discovered in the three directions $D_{u}$, $D_{d}$, and $D_{r}$, respectively; however, only $v_{9}$ is considered a new state.

The third iteration similarly starts from the newly discovered states, $v_{3}$, $v_{6}$, and $v_{9}$. In the same way, considering all four directions from each state and filtering out repeated states, four new states are identified: $v_{4}$, $v_{7}$, $v_{10}$, and $v_{13}$.

EvoCreeper works in an iterative evolutionary manner to discover new states and events in the app. As mentioned, a cloud-based app is considered in this pilot example. Hence, there is no expectation that the app will have a finite number of states. Consequently, our proposed alternative stopping criterion could be useful here. The creeper algorithm will stop after a certain number of iterations or when no new states are discovered.

\section{Empirical Evaluation}\label{Evaluation}

To assess the effectiveness of our model generation strategy, we conducted a case study on four smart TV apps. During this evaluation, we attempted to address the following research questions (RQs):

\begin{itemize}
\item RQ1. Is EvoCreeper capable of exploring and identifying the states and transitions of a given smart TV app accurately relative to manual exploration? 
\item RQ2. To what extent is the generated model complete in terms of the numbers of states and edges? Is the created graph valid? 
\item RQ3. What is the performance of EvoCreeper compared to that of manual exploration for a given smart TV app?   
\end{itemize}

\subsection{Experimental Objects} 

Research in the area of smart TV apps is in an early stage. More time may be needed for developers to create and publish smart TV apps, as it is a new development environment. As a result, not many apps and repositories are available for benchmarking. Tizen does maintain a page with several simple apps and examples\footnote{https://bit.ly/2qC5ncS}. However, most of the provided samples are simple apps with few states. To demonstrate the effectiveness of our strategy, we chose four different Tizen smart TV apps of different sizes from GitHub. These apps are from different domains and have varying numbers of states. Table \ref{AppObjectsForExperiments} shows the name and source of each app.

\begin{table*}
\centering
\caption{APPS USED IN THE CASE STUDY}\label{AppObjectsForExperiments}
\scriptsize
\begin{tabular}{|c|l|>{\raggedright}p{10cm}|}
\hline 
ID & App & Source\tabularnewline
\hline 
\hline 
1 & CineMup & https://github.com/daliife/Cinemup\tabularnewline
\hline 
2 & ChessLab TV & https://github.com/PabloEzequiel/Tizen/tree/master/ChessLabTV\tabularnewline
\hline 
3 & MonitorDeLoterias & https://github.com/brunohpmarques/monitordeloterias-tizentv\tabularnewline
\hline 
4 & Memory game & https://github.com/wissalKhalfi/Brain-Up-{}-{}-Tizen-smart-TV/tree/master/Memoryyyyyy\tabularnewline
\hline 
\end{tabular}

\end{table*}

The chosen apps are CineMup, ChessLab TV, Monitor de Loterias, and the game Memory. CineMup is an app for searching for various types of movies and TV shows. The app categorizes movies and shows and presents relevant data for each along with a trailer. ChessLab TV is an app for improving one’s personal skills in chess by presenting tactics and solutions to puzzles. Monitor de Loterias is a Spanish lottery gaming app. Memory is a gaming app for memory-based puzzles.

\subsection{Experimental Procedure}\label{ExperimentalDesign}

To address RQ1 and RQ3, we conducted a set of experiments. The goal was to compare the models of the benchmark apps created by EvoCreeper with models created through manual exploration. For the experiments, we adopted the following set-up.

We instructed four groups of 60 students from the advanced software engineering study program to download a software package consisting of a portion of Tizen Studio that allows its emulator to be run on a desktop or notebook computer. We extended the remote control module in the emulator to record the exploration history in each smart TV app, consisting of the source UI elements, the particular remote control keys pressed, the target elements and the corresponding timestamps. Each member of the group was assigned one app to explore and was instructed to export the exploration history log when finished. The stopping criterion was "the participant considers all parts of the app to have been explored".

We then processed the recorded logs to construct the app models. For each model, we analyzed several properties: the time needed to create the model (to explore the smart TV app to the extent captured by the model), the number of nodes in the model, the number of unique nodes in the model (thus enabling the determination of the level of duplicity of the nodes in the model) and the number of edges.

\subsection{Model-Checking Algorithm}

To address RQ2, we developed an algorithm to check the consistency of the graphs generated by EvoCreeper. Each generated model graph was validated with Algorithm \ref{VerificationAlgo}.

\begin{algorithm}

\KwIn{$G$}  
\KwOut{list of issue tokens $I$}

\If {$G=\emptyset$} { $I\leftarrow I\cup empty\_model$}

\If {$n_{s}$ not exist} {$I\leftarrow I\cup no\_start\_node\_defined$}

\ForEach {$n\in N\setminus\{n_{s}\}$}{

\If {$deg^{\text{-}}(n)=0$} {$I\leftarrow I\cup node\_n\_has\_not\_incomming\_edge$}}

\If {there is not at least one $n\in N\setminus\{n_{s}\}$
that $deg^{\text{+}}(n)=0$} {$I\leftarrow I\cup model\_has\_not\_end\_node$}

\ForEach {$n_{1}\in N\setminus N_{e}$}{

\If {there does not exist a path from $n_{1}$ to
any node from $N_{e}$} {$I\leftarrow I\cup end\_node\_for\_n_{1}\_not\_reachable$}}

\ForEach {$n\in N\setminus\{n_{s}\}$}{

\If {there does not exist a path from $n_{s}$ to
$n$} {$I\leftarrow I\cup node\_n\_not\_reachable\_from\_start$}}

\caption{\label{VerificationAlgo} The app model validation algorithm }

\end{algorithm}

As can be seen from Algorithm \ref{VerificationAlgo}, during this validation, several steps are performed. The model must not be empty, its starting node must be defined, and all nodes of the model except the starting node must have at least one incoming edge. Furthermore, at least one end node must be present in the model. From each node, it must be possible to reach an end node. Finally, each of the nodes must be reachable from the starting node.

\subsection{Results}

As previously mentioned, we created a set of experiments using the Tizen SDK environment to assess the effectiveness of our EvoCreeper strategy. Additionally, we developed an algorithm for automatically verifying the correctness of the models thus generated. The goal of these experiments was to address three main RQs. The following subsections answer these RQs.

\subsubsection{RQ1. State detection comparison of EvoCreeper and manual exploration}

\begin{figure*}[tbph]
	\centering
	\begin{subfigure}[b]{0.24\textwidth}
		\includegraphics[width=\textwidth]					{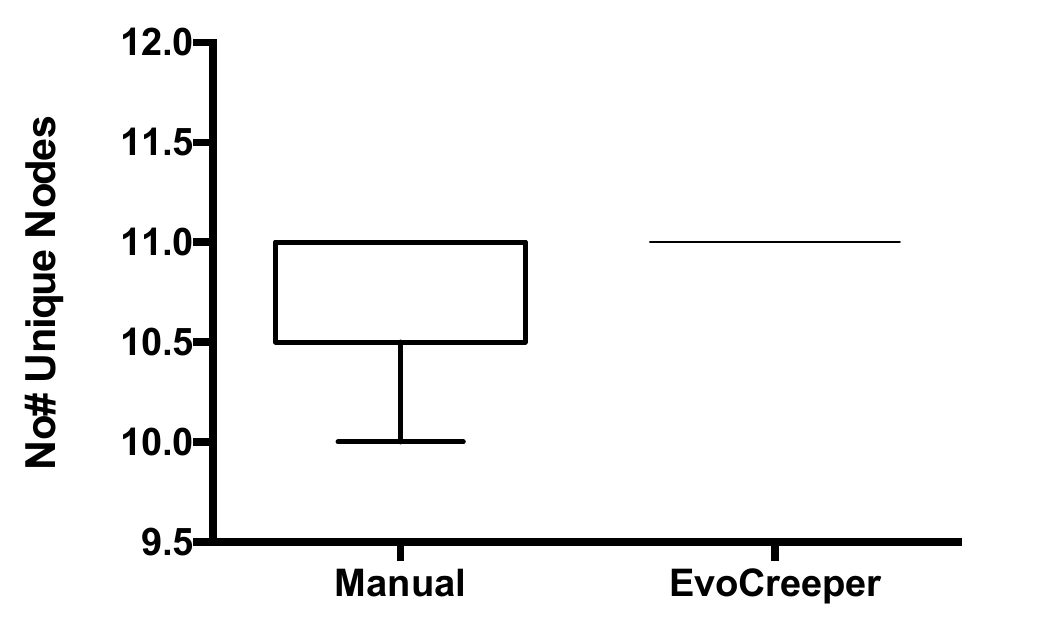}
                \vspace{-0.3cm}
		\caption{ChessLab \centering }
		\label{ChessLabNodeNumber}
    \end{subfigure}%
         \hfill
    \begin{subfigure}[b]{0.24\textwidth}
    	\includegraphics[width=\textwidth]					{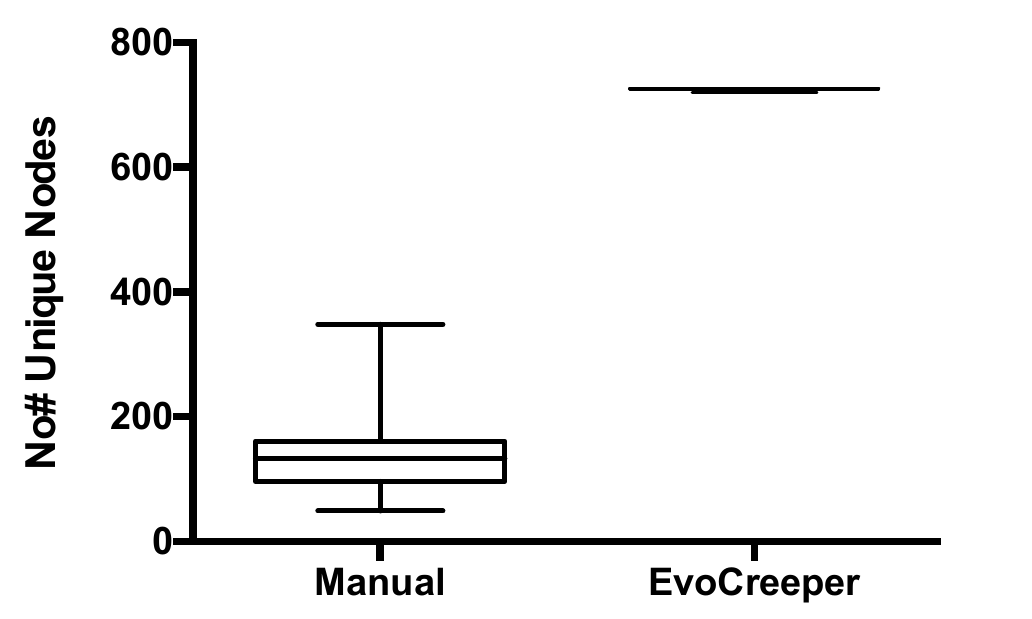}
                \vspace{-0.3cm}
        \caption{CineMup \centering }
        \label{CineMupNodeNumber}
          \end{subfigure}
             \hfill
             \begin{subfigure}[b]{0.24\textwidth}
    	\includegraphics[width=\textwidth]					{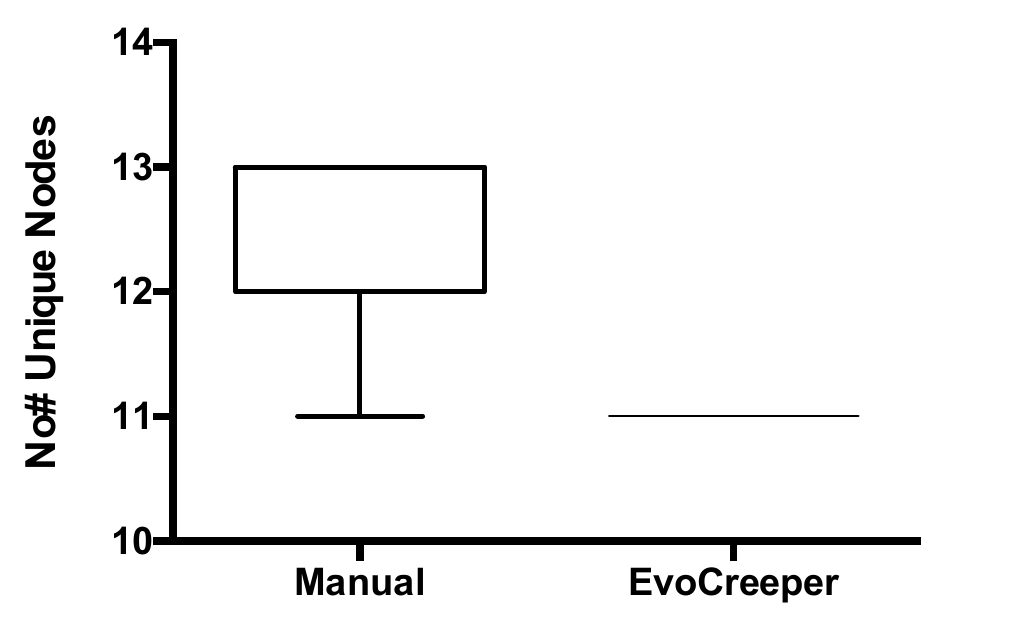}
                \vspace{-0.3cm}
        \caption{Memory game  \centering }
        \label{MemorygameNodeNumber}
          \end{subfigure}
             \hfill
    \begin{subfigure}[b]{0.24\textwidth}
    	\includegraphics[width=\textwidth]					{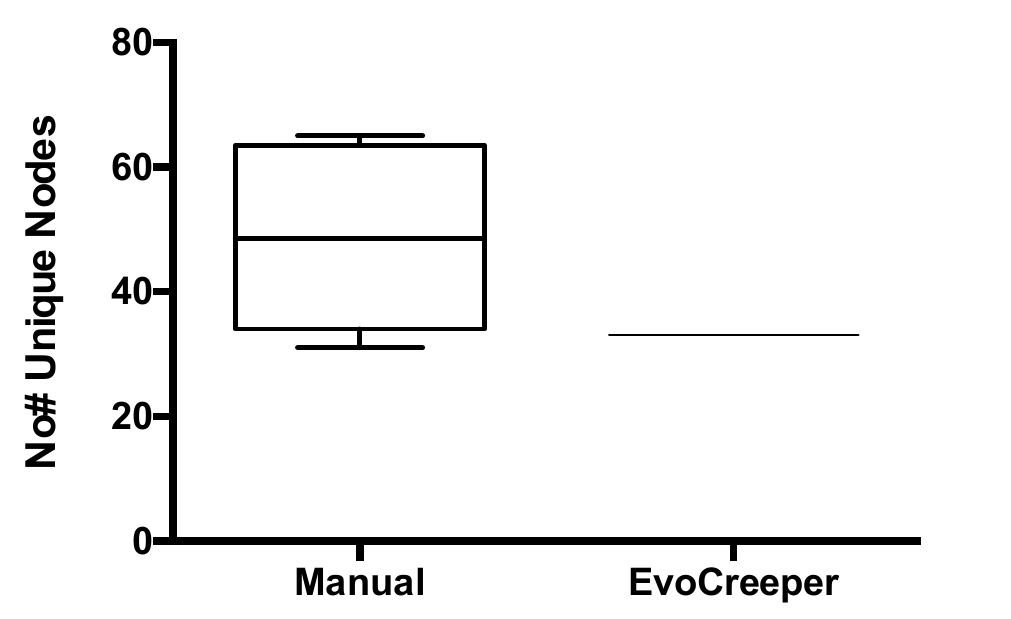}
                \vspace{-0.3cm}
        \caption{MonitorDeLoterias \centering }
        \label{MonitorDeLoteriasNodeNumber}
    \end{subfigure}
\caption{Comparing the number of unique nodes detected by EvoCreeper with the manual exploration}
\label{NumberOfUniqueNodeBoxPlot}
\end{figure*}

\begin{figure*}[tbph]
	\centering
	\begin{subfigure}[b]{0.24\textwidth}
		\includegraphics[width=\textwidth]					{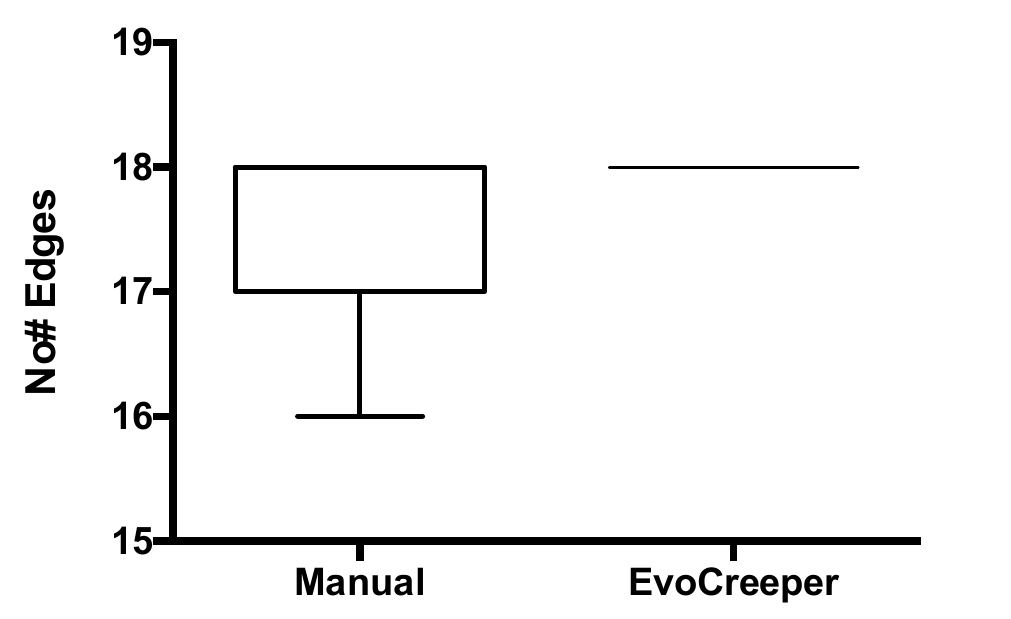}
                \vspace{-0.3cm}
		\caption{ChessLab \centering }
		\label{ChessLabEdgeNumber}
    \end{subfigure}%
         \hfill
    \begin{subfigure}[b]{0.24\textwidth}
    	\includegraphics[width=\textwidth]					{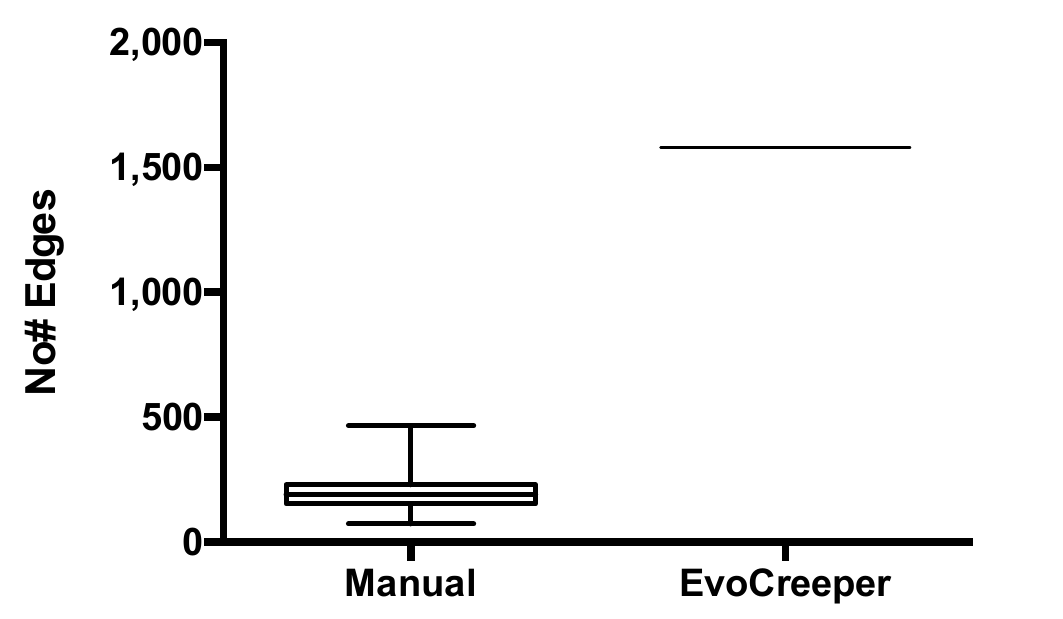}
                \vspace{-0.3cm}
        \caption{CineMup \centering }
        \label{CineMupEdgeNumber}
          \end{subfigure}
             \hfill
             \begin{subfigure}[b]{0.24\textwidth}
    	\includegraphics[width=\textwidth]					{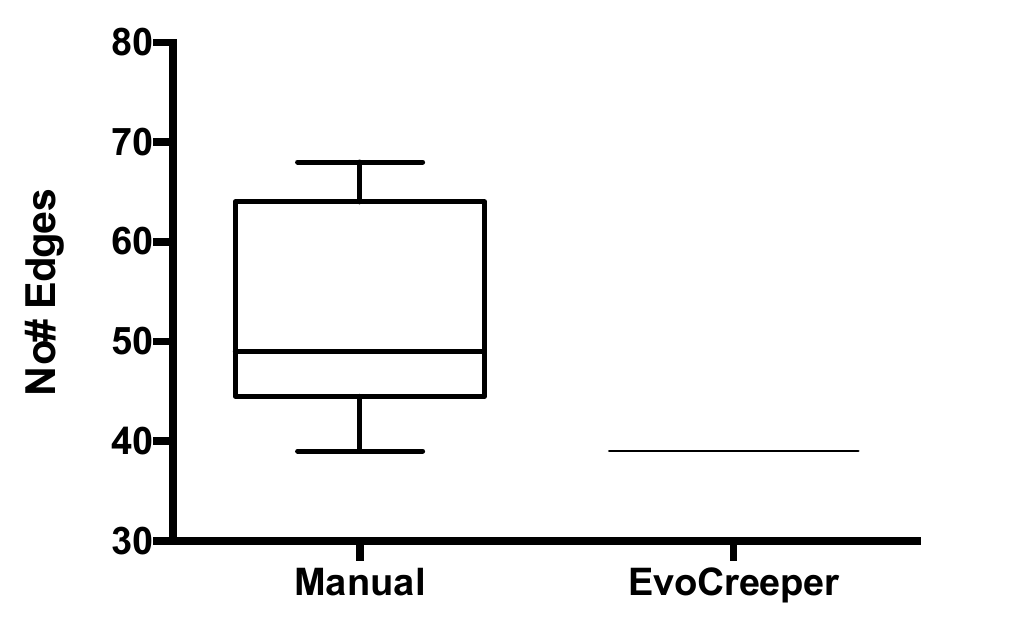}
                \vspace{-0.3cm}
        \caption{Memory game  \centering }
        \label{MemorygameEdgeNumber}
          \end{subfigure}
             \hfill
    \begin{subfigure}[b]{0.24\textwidth}
    	\includegraphics[width=\textwidth]					{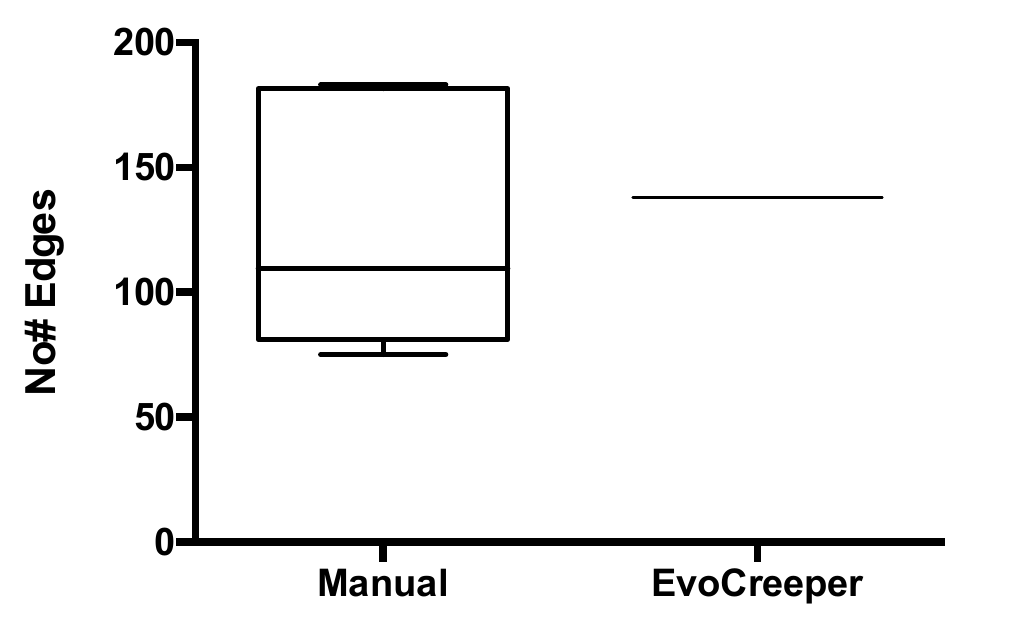}
                \vspace{-0.3cm}
        \caption{MonitorDeLoterias \centering }
        \label{MonitorDeLoteriasEdgeNumber}
    \end{subfigure}
\caption{Comparing the number of Edges detected by EvoCreeper with the manual exploration}
\label{NumberOfUniqueEdgesBoxPlot}
\end{figure*}

To answer this RQ, we compared the numbers of unique states and edges detected by our strategy with those identified through the manual exploration of the apps. The experimental procedure is described in Section \ref{ExperimentalDesign}. Figures \ref{NumberOfUniqueNodeBoxPlot} and \ref{NumberOfUniqueEdgesBoxPlot} show the box plots of the results for comparison.

Due to the subjective and nondeterministic nature of the results obtained by each participant, we compare the results using box plots to ensure fair comparisons and provide more details. It is clear from the figures that in contrast to manual exploration, our strategy produces deterministic results. We can observe from the results that for large apps, our strategy can produce more accurate results than manual exploration can. Regarding the numbers of unique states in Figure \ref{NumberOfUniqueNodeBoxPlot}, it is clear that our strategy can generate better results than can be obtained through manual exploration in most cases. In fact, we notice that manual exploration can also be effective for some members of the exploration group, but only for small apps with few states. For example, the best result achieved among the participants for the number of nodes is much lower than that achieved by our strategy for the CineMup app, as seen in Figure \ref{CineMupNodeNumber}. A similar situation is also clearly observed in Figure \ref{CineMupEdgeNumber} for the number of edges discovered. For a small application such as ChessLab TV, as seen in Figures \ref{ChessLabNodeNumber} and \ref{ChessLabEdgeNumber}, manual exploration can reveal as many states and edges as automated exploration can in some cases. In fact, the success of manual exploration depends on the participant’s knowledge and experience with the app. While our strategy explored each app without any prior knowledge about its operation, the participants were able to explore the apps several times before the recording the exploration logs.

We note that for some apps, the results of our strategy are missing a few states relative to the results of manual exploration. For instance, the output of our strategy is missing two states and their related edges for the Memory game app, as seen in Figures \ref{MemorygameNodeNumber} and \ref{MemorygameEdgeNumber}. A few more nodes and edges are missing relative to those found through manual exploration for the Monitor de Loterias app, as seen in Figures \ref{MonitorDeLoteriasNodeNumber} and \ref{MonitorDeLoteriasEdgeNumber}. This situation arises due to the nature of these two apps. Both apps require some manual text entry at some stage during operation. This explains the deficiency of our strategy in these cases. In fact, even in the reverse engineering of desktop and mobile apps, such situations are a critical concern because the required text is specific to the particular app and may change depending on the app itself. This difficulty can be addressed by defining the text to be entered into each specific app.

Overall, manual exploration is feasible for model generation for small apps; however, our strategy is the best choice for small, medium and large apps when there are many states to be explored. Additionally, manual exploration is neither a realistic nor a practical method of model creation.

\subsubsection{RQ2. Graph verification and correctness}

To verify the correctness of the graphs generated by our strategy, we developed our automated verification method presented in Algorithm \ref{VerificationAlgo}. This algorithm checks a graph for incomplete and invalid paths and nodes. In fact, it simply checks the validity of a directed graph generated by our strategy. Table \ref{VerificationTableResults} shows the results of this verification for each app used in our experiments.

\begin{table}
\centering
\caption{GRAPH VERIFICATION RESULTS BY ALGORITHM \ref{VerificationAlgo}  \label{VerificationTableResults}}
\begin{tabular}{|c|l|l|l|}
\hline 
ID & App & No. of not reachable path & Result\tabularnewline
\hline 
\hline 
1 & CineMup & 0 & Pass\tabularnewline
\hline 
2 & ChessLab TV  & 0 & Pass\tabularnewline
\hline 
3 & MonitorDeLoterias & 0 & Pass\tabularnewline
\hline 
4 & Memory game & 0 & Pass\tabularnewline
\hline 
\end{tabular}
\end{table}

We can clearly see that for all apps, the graphs were generated correctly, and the algorithm could not find any invalid paths with a missing state or edge.

\subsubsection{RQ3. Performance comparison of EvoCreeper with manual}

\begin{figure*}[tbph]
	\centering
	\begin{subfigure}[b]{0.35\textwidth}
		\includegraphics[width=\textwidth]					{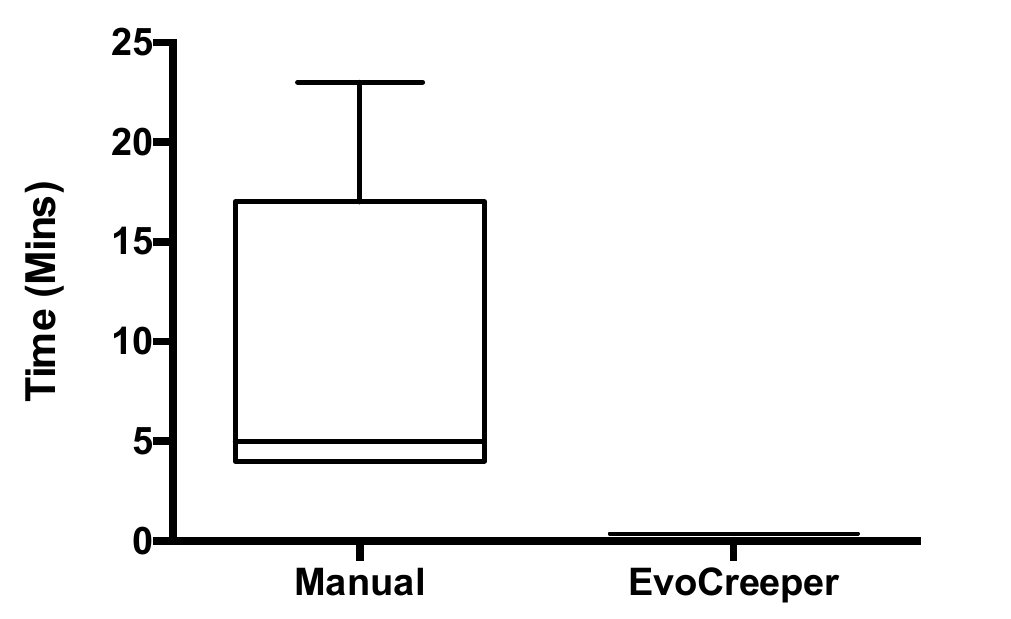}
                \vspace{-0.3cm}
		\caption{ChessLab \centering }
		\label{ChessLabTime}
    \end{subfigure}%
         \hfill
    \begin{subfigure}[b]{0.35\textwidth}
    	\includegraphics[width=\textwidth]					{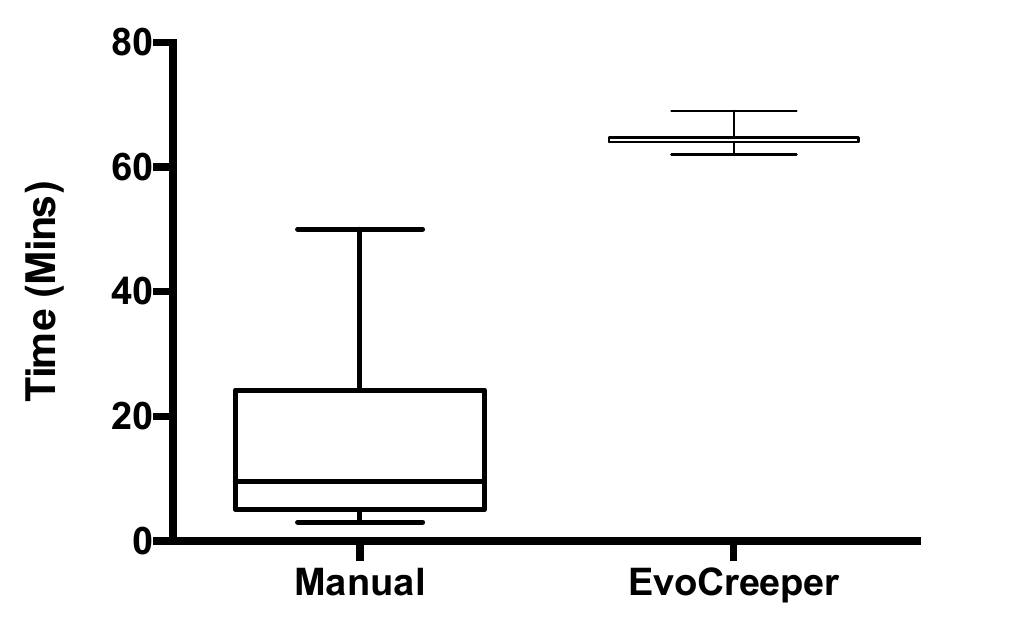}
                \vspace{-0.3cm}
        \caption{CineMup \centering }
        \label{CineMupTime}
          \end{subfigure}
             \hfill
             \begin{subfigure}[b]{0.35\textwidth}
    	\includegraphics[width=\textwidth]					{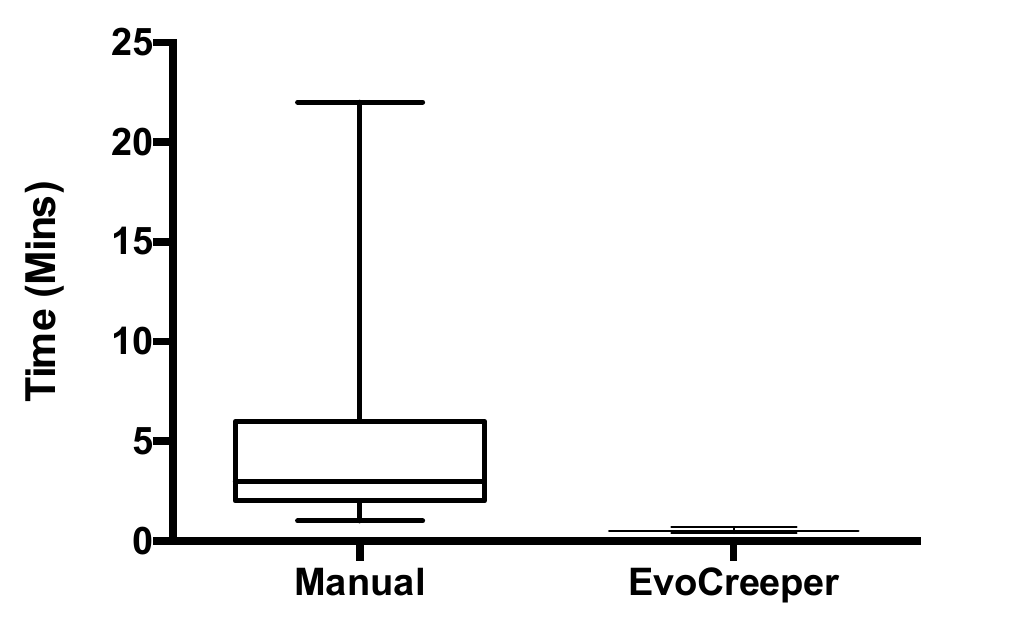}
                \vspace{-0.3cm}
        \caption{Memory game  \centering }
        \label{MemorygameTime}
          \end{subfigure}
             \hfill
    \begin{subfigure}[b]{0.35\textwidth}
    	\includegraphics[width=\textwidth]					{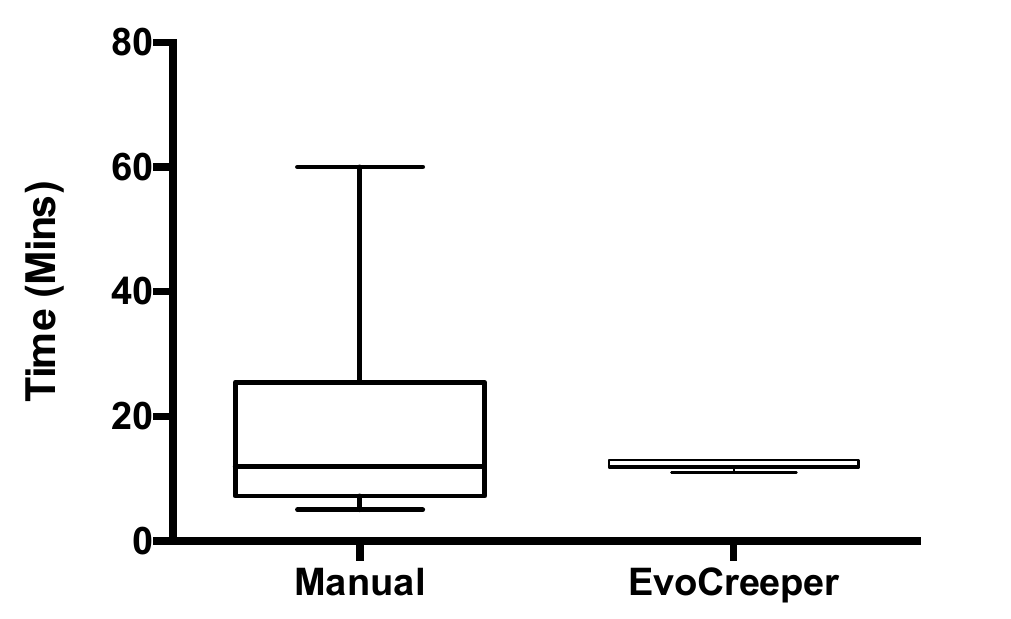}
                \vspace{-0.3cm}
        \caption{MonitorDeLoterias \centering }
        \label{MonitorDeLoteriasTime}
    \end{subfigure}
\caption{Comparing the exploration time by EvoCreeper with the manual exploration}
\label{TimeNodeBoxPlot}
\end{figure*}

Another critical objective of this study was to assess the performance of our strategy compared to that of manual exploration. As mentioned previously, we measured the time required for manual exploration for each participant individually by calculating the difference between the start and end times of exploration from the log. Figure \ref{TimeNodeBoxPlot} shows the corresponding results for each app.

It is clear from Figure \ref{TimeNodeBoxPlot} that the exploration times varied among the participants. As mentioned previously, the exploration time may change from one participant to another depending on the usability of the app and that participant’s knowledge of the app. To ensure fair comparisons and obtain more details about the results, we used box plots to present the findings. While the participants’ times vary, the corresponding range of variation is much lower for our strategy, as can be clearly seen from the low interquartile ranges in Figure \ref{TimeNodeBoxPlot}.

It is clear from Figure \ref{MonitorDeLoteriasTime} that for a medium-size app, the time required by our strategy is less than the median time required by the participants. For some of the other apps, as seen in Figures \ref{ChessLabTime} and \ref{MemorygameTime}, our strategy even takes less time than the best time recorded by any participant.

For large apps such as CineMup, for which the results are shown in Figure \ref{CineMupTime}, our strategy takes more time due to the large number of states to explore. In fact, this time is much higher than the worst time taken for manual exploration among the participants. However, this time is reasonable with respect to the accuracy and completeness of the generated graph. We note from Figure \ref{CineMupEdgeNumber} that the number of unique nodes found by our EvoCreeper strategy is much higher than the best result achieved by any participant. Specifically, our strategy found 727 unique states in approximately 64 minutes of exploration time, whereas the best group member found 348 unique states in approximately 50 minutes of exploration time.

\section{Threats to Validity}\label{Threats}

Like any empirical and experimental study, the validity of the evaluation and case studies presented in this paper is not without any threats. In fact, there are a few factors that may affect the validity of the presented results. We have attempted to eliminate each of these factors, as described below.

The comparison with the results of manual exploration of the apps could pose an internal threat to validity. As mentioned earlier, research in the area of smart TV apps is in an early stage. Hence, we could not find another reverse engineering tool against which to compare our algorithm. Here, the manual exploration is a well-known approach that is normally used for different testing purposes (e.g., \cite{Raappana2016, Hellmann2011, Bures2018}). However, manual exploration is subjective and could be affected by participants’ biases. We mitigated this threat by considering many participants in our experiment. We also presented the results as box plots to visualize the best, worst, and median results.

A possible threat can be identified regarding the simulation of the smart TV apps on an emulator instead of executing them on a real device. However, for app exploration, there is no significant difference between the actual device and the emulator. Here, we were simply creating a model, and we do not expect any differences in the model between the two cases because it is a platform-independent model.

Another threat could be posed with regard to the recruitment of students for the experiment. When an app is given to an experienced professional software engineer or smart TV app developer, for instance, the generated model could be completed in a slightly shorter time. As mentioned previously in the discussion of the results, the usability of an app and prior knowledge about it could affect the exploration time and the model completeness. We attempted to mitigate these factors by illustrating the app workflow to each student. Additionally, we provided sufficient time for the students to learn and explore the apps before the experiment. Each student was also able to run the experiment several times and report the best time achieved. However, many other human factors may affect the results, and indeed, eliminating these factors is an essential aim of our strategy. Automating the model generation process will eliminate the influence of human factors and, consequently, the subjectivity of the results. Even with our mitigation strategies, the results show that our automated black-box strategy is superior. For a user who does not have any information about the app being explored, we would expect worse results than those presented in this paper.

Another threat is related to the fact that some smart TV apps use different remote control keys other than \textit{Left}, \textit{Right}, \textit{Up}, \textit{Down}, \textit{OK} and \textit{Back}. In some cases, this could influence the completeness of the model. However, this fact does not affect the reliability of the present experiment because the participants were restricted by the logging mechanism in the emulator to creating the manual models of the apps by using the \textit{Left}, \textit{Right}, \textit{Up}, \textit{Down}, \textit{OK} and \textit{Back} keys only. The extra keys on a remote control device are typically app-specific keys and may not strongly affect the general model. Instead, they are likely to affect the model completeness because some states and edges could be missing.

Finally, for apps that depend on cloud content, the model may change over time. This could lead to differences in the model between one time and another. Hence, two different participants may generate two different but valid models for the same app on two different days. We mitigated this threat by ensuring that the participants created their models at the same time. In fact, this threat reflects another benefit of our strategy. Using EvoCreeper will enable the user to create up-to-date models for apps at different points in time without requiring manual effort.

\section{Conclusion and Future Work}\label{Conclusion}

In this paper, we have presented our strategy for automatically reverse engineering smart TV apps considering interaction via a remote control device. This strategy involves navigating a given smart TV app to explore its UI states without knowing the internal structure of the app code (i.e., it is a black-box strategy). The strategy can extensively explore the states and transitions in a given app and then generate a directed graph model. We have implemented our strategy in a tool called EvoCreeper that works with the emulator in the Tizen SDK framework. We have evaluated our strategy using four medium-size and large apps. The evaluation results demonstrate the effectiveness and good performance of our strategy. The strategy can be used to detect states that would go undetected through manual efforts. The generated models could be used in various stages of app development for quality improvement, testing, identifying missing states, assessing the user experience and understanding of a given app by providing a visualization of the states and transitions, and even more.

There are many opportunities for future research. An immediate step forward is to use the models generated via the presented strategy to test various apps and identify new faults. Additionally, we are planning to use this strategy for the smoke testing of smart TV applications. Part of this process will be to design an algorithm for generating path-based test cases from such a generated model. Another important direction of research that we are planning for the future is to investigate model generation for gaze-based interaction with smart TV apps.

\bibliographystyle{IEEEtran}
\bibliography{sample-bibliography}

\begin{IEEEbiography}[{\includegraphics[width=1in,height=1.25in,clip,keepaspectratio]{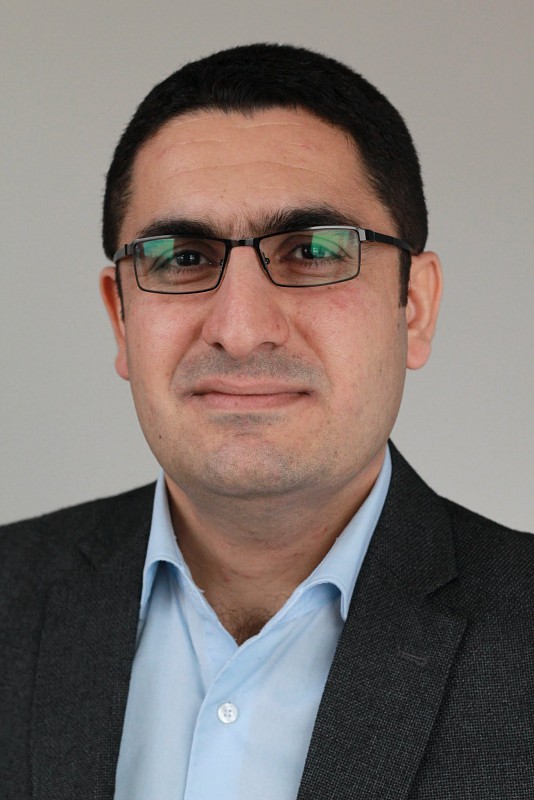}}]{Bestoun S. Ahmed}
obtained his B.Sc. degree in Electrical and Electronic Engineering from the Salahaddin University-Erbil in 2004, his M.Sc. degree from University Putra Malaysia (UPM) in 2009, and his Ph.D. degree from University Sains Malaysia (USM), Software Engineering, in 2012. He spent one year doing his post doctoral research in the Swiss AI Lab IDSIA, Switzerland. Currently, he is a senior lecturer at the department of mathematics and computer science, Karlstad University, Sweden and also a part time assistant professor at the department of computer science, Czech Technical University in Prague. His main research interest include Combinatorial Testing, Search Based Software Testing (SBST), Applied Soft Computing, and quality assurance of smart devices and IoT systems.
\end{IEEEbiography}

\begin{IEEEbiography}[{\includegraphics[width=1in,height=1.25in,clip,keepaspectratio]{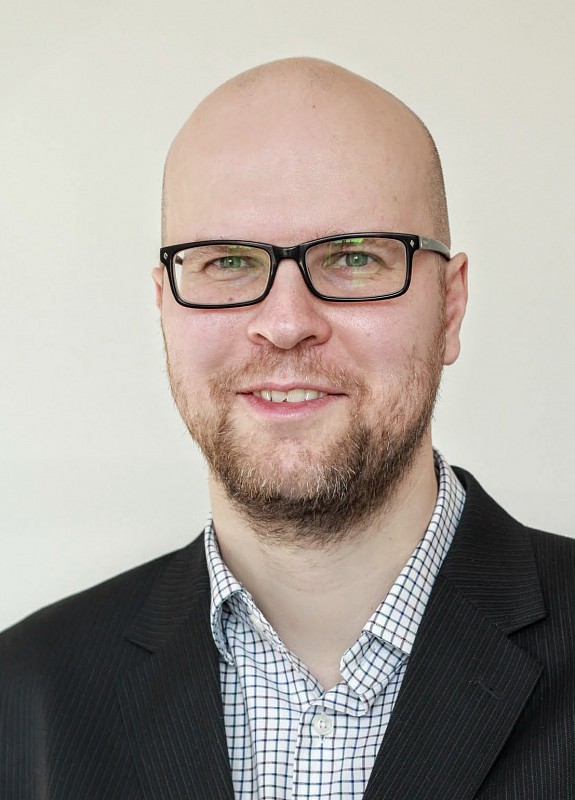}}]{Miroslav Bures}
received his Ph.D. at Czech Technical University in Prague, Faculty of Electrical Engineering, where he currently works as a researcher and a senior lecturer in software testing and quality assurance. His research interests are model-based testing (process and work flow testing, data consistency testing) efficiency of test automation, and quality assurance methods for Internet of Things solutions, reflecting specifics of this technology. He is a member of Czech chapter of the ACM, CaSTB, and ISTQB Academia work group.
\end{IEEEbiography}

\end{document}